# In situ characterisation of slip bands behaviour in ferrite under mechanical loading

Abdalrhaman Koko[1a,b]

[a] National Physical Laboratory, Hampton Road, Teddington TW11 0LW, UK

[b] Department of Materials, University of Oxford, Oxford OX1 3PH, United Kingdom

## Abstract

This study investigates the behaviour of slip bands, terminated mid-grain in the ferrite grains of age-hardened stainless steel, under different conditions to understand their dislocation activity and response to varying loads. The full Nye lattice curvature tensor was measured in situ using high-resolution electron backscatter diffraction (HR-EBSD) to estimate the total geometrically necessary dislocation density and individual mobile and immobile dislocations activity at the slip band scale. We found that slip bands primarily consist of edge dislocations, marked by a 'blooming zone' at its tip, indicating significant shear deformation and loss of mobile dislocations. The blooming zone expands as the load increases while the slip band thickness remains constant. Coupled with the correlation between mobile and immobile dislocations within the loaded slip bands observed in situ, we observed that the dislocation activities at the slip band were mainly from immobile edge dislocations to maintain the geometrical distortion of the slip band.

---

[1] E-mail address: abdo.koko@npl.co.uk

# 1. Introduction

Dislocations play a crucial role in the plastic deformation of crystalline materials. These line defects represent the misalignment of the atoms within the crystal lattice. Dislocations can be classified into different types, and their behaviour and distribution affect the mechanical properties of materials. This study focused on two types of dislocations within loaded slip bands: mobile (or glissile) and immobile (or sessile) dislocations. Mobile dislocations, capable of moving within the crystal lattice under an applied load, directly contribute to plastic deformation. These are associated with slip bands and regions where plastic deformation occurs [1–3]. On the other hand, immobile dislocations, which are pinned or locked due to interactions with other dislocations or obstacles in the lattice, contribute to work hardening and strengthening mechanisms in the material. Although they may not directly cause plastic deformation like mobile dislocations, they are necessary to accommodate the strain caused by plastic deformation, particularly in the context of plastic bending. They arise because of the internal plastic strain gradients and help to relieve the strain by creating a geometrically necessary dislocation structure that accommodates the curvature of bent crystal planes [4,5]. Both mobile and immobile dislocations arise owing to the incompatibility of plastic strain, and understanding the relationship between mobile and immobile dislocations is crucial for understanding the mechanisms underlying the plastic deformation processes [1,2].

To enable this understanding, we first must accurately measure the dislocation density. High-resolution electron Backscatter Diffraction (HR-EBSD) analysis has emerged as a powerful technique for investigating dislocations and their effects on materials. It allows the mapping of lattice strain, dislocation density, and dislocation structures at a high spatial resolution [6,7]. HR-EBSD can provide valuable insights into the behaviour of dislocations in various materials, including metals, alloys, and geological minerals [6,8]. One important aspect of HR-EBSD analysis is the ability to study strain localisation and distribution near dislocations. It has been observed that strain localisation is not limited to regions near grain boundaries or triple junctions. For example, in polycrystalline copper, HR-EBSD analysis has shown that different grain orientations exhibit distinct dislocation structures that can influence the strain distribution [8].

The spatial resolution of HR-EBSD is determined by the electron interaction volume, which typically ranges from several tens to hundreds of nanometres in bulk materials. Although the spatial resolution of HR-EBSD has been a limitation for studying the strain fields associated with individual dislocations, statistical analysis suggests that there is sufficient spatial resolution to probe the lattice distortion near dislocations [7]. In addition, the geometrically necessary dislocation (GND) density calculation based on the local kernel disorientation from the HR-EBSD elastic lattice rotation gradient exhibits a lower noise level than that based on the local kernel average misorientation (KAM) from the Hough-EBSD Euler angles [9]. This was due to the increased angular resolution when calculating the lattice rotation. However, the estimated GND distribution negligibly depends on the reference pattern selection because only a lattice distortion gradient is required [10].

HR-EBSD can characterise the roles of different dislocation types during crystal plastic deformation and map heterogeneous internal stress fields associated with specific deformation mechanisms or changes in temperature, pressure, or stress [6]. This makes HR-EBSD a reliable tool for measuring dislocation density and enables a deeper understanding of the mechanisms underlying plastic deformation processes. This was demonstrated in our previous work [11], where HR-EBSD was used to examine the local elastic field at the tips of mechanically loaded intragranular slip bands in the ferrite phase of age-hardened duplex stainless steel. There, we examined the elastic deformation fields at the tip of slip bands, and by mathematically integrating the surface elastic strain field to calculate in-plane and out-of-plane surface displacements [12], we parameterised the elastic deformation fields by calculating the strain energy release rate and relating it to the stress intensity factors using finite element analysis, which used the displacement field as the boundary condition. Our study concluded that the stress intensity factors ahead of the slip band, measured under load, were found to be directly affected by the magnitude of loading and the inclination angle of the slip band to the observed surface, which provided valuable insights into the behaviour of intragranular slip bands in ferrite.

Thus, the current analysis aims to study the relationship between mobile and immobile dislocations within loaded slip bands that are terminated mid-grain and were observed – in situ – using HR-EBSD, which has the necessary spatial resolution and sensitivity. We will also look at the deformation field induced by the slip band as it propagates inside the grain and the interplay between the stress and GND density that enables slip band intragranular propagation.

## 2. Methodology

### 2.1. Sample preparation and mechanical testing

The Zeron 100 duplex stainless steel sample, which consisted of austenite (face-centred cubic) and ferrite (body-centred cubic) phases, was hardened by aging in air at 475°C for 100 h. This process aims to harden ferrite grains while maintaining the mechanical structure of austenite [13,14]. Subsequently, the sample was polished to a mirror-like finish and subjected to deformation using 3-point bending. Bending was performed using a 2 kN Deben® 70° pre-tilted loading stage with a Carl Zeiss Merlin FEG-SEM. The deformation was paused at specific displacement increments to collect electron backscatter diffraction (EBSD) maps. The examination locations were selected using secondary electrons (SE) and back- and fore-scatter electron imaging techniques known as virtual forward-scattered electron detectors (VFSDs) [15].

The main objective was to identify slip bands that terminated mid-grain or intragranular slip bands, which display a localised field without interactions with other stress concentrators such as additional slip bands, grain boundaries, or phase boundaries. This will simplify the analysis and avoid contributions from other stress localisers. Thus, the selected grains were between the stress's neutral axis and the tensile stress region of the 3-point bending test. See the Supplementary Figures for more details, and further details of this experiment with more focus on elastic strains can be found in [11].

### 2.2. HR-EBSD

In a conventional EBSD setup, an in situ mapping technique was employed to collect a grid of 800 × 600 electron backscatter patterns (EBSPs). A Bruker eFlash CCD camera was used with beam conditions of 10 nA/20 kV, exposure time of 100 ms per pattern, and step size of 75 nm. From the analysis of Ruggles et al. [16] and Suchandrima et al. [17], any step size between 500 and 50 nm should be adequate to estimate the geometrically necessary dislocation density correctly.

Within each grain, the EBSPs were divided into 30 regions and cross-correlated with a reference pattern (EBSP$_0$). This process allowed for the measurement of shifts and estimation of the elastic deformation gradient tensor, using the HR-EBSD method [16] as implemented in CorssCourt®. To minimise the effect of the reference pattern, an objective process was used to select the least-strained reference pattern by considering the empirical relationship between the mean angular error (MAE) and pattern quality (PH). The reference pattern with the highest PH and lowest MAE was identified [10]. In addition, to improve the calculation accuracy, the analysis was performed iteratively, and in the second iteration, the EBSPs were remapped to the estimated the EBSP orientation which was obtained in the first iteration [17]. This procedure aims to minimise errors and enhance the reliability of the results, achieving an average mean angular error of $2.7 \times 10^{-4}$.

## 2.3. Geometrically necessary dislocation (GND) density estimation

Plastic deformation is accommodated by dislocation behaviour, and an effective way to analyse intragranular misorientation can be achieved by analysing the residual dislocations accumulated during hardening (i.e. GND density). This was achieved using the Nye [4]–Kröner [18]–Bilby [19] framework for EBSD data to estimate the GND content (i.e. density and arrangement), which was linked to the slight local lattice curvature caused by stress incompatibility [20,21], as shown in equation (1).

$$\alpha_{ij} = \underbrace{\frac{1}{V} \int_L b_i\, t_j\, ds}_{\text{Avg. Dislocations Properties}} = \underbrace{\frac{1}{V} \sum_{d=1}^{D} b_i^d \int_l t_j^d\, ds^d}_{\text{Retain Dislocations Properties}} = \sum_{d=1}^{D} \rho^d * b_i^d \otimes t_j^d \qquad 1$$

$$\therefore \rho^d = \frac{1}{V} \int_L ds^d = \frac{l^d}{V} = \rho_{GNDs}^d + \underbrace{\rho_{SSDs}^d}_{\sum = 0} = \rho_{GNDs}^d \qquad 2$$

where $\alpha_{ij}$ is Nye's dislocation tensor, $L$ is the total length of the dislocation in the reference volume ($V$), $l$ is the length of the dislocation type ($d$) with the Burgers vector $b_i$, $ds$ is the length increment along the dislocation line, $t_j$ is the line vector, $\rho^d$ is the GND density of the dislocation type ($d$), and D is the all present dislocation types (e.g. 18 for FCC with 12 edges and 6 screw dislocations [5]).

Using this approach, individual dislocation contents can also be resolved without serial sectioning or knowledge of the sample deformation history [22,23]. Although the net geometric effect of statistically stored dislocations (SSDs) is cancelled by adjacent opposite-sign dislocations within the Burgers circuit (i.e. the net Burgers vector equals zero), as shown in equation (2); extreme care must be exercised when deciding the interaction volume scale (or the measurement step size) for non-redundant-dislocation measurements. This is because a step size that is too large can lead to missing essential dislocations as one portion is averaged with another portion. Still, a step size that is too small can render the dislocation density no longer continuously distributed [5,20,24].

Moreover, considering a continuous displacement field $u(x)$ defined only in an infinitesimal neighbourhood $\left(1 \gg \left|\frac{\partial u}{\partial x}\right|\right)$ of the position $x$ around the closed curve (C); the relationship between the local lattice curvature (equation 3) and local dislocation networks (equation 4), which is written as the net Burgers' vector (B) over the surface ($s$) can be obtained [4,20,21].

$$\oint_c du = \oint_c \beta dx = \oiint_s \mathrm{curl}(\beta)\, ds = \oiint_s \mathrm{curl}(\beta^{\mathrm{Plastic}} + \beta^{\mathrm{Elastic}})\, ds = 0 \qquad 3$$

$$\mathrm{B} = -\oint_c \beta^{\mathrm{P}}\, dx = -\oiint_s \mathrm{curl}(\beta^{\mathrm{P}})\, ds = \oiint_s \alpha_{ij}\, ds \qquad 4$$

$$\alpha_{ij} = -\mathrm{curl}(\beta^{\mathrm{P}}) = \mathrm{curl}(\beta^{\mathrm{E}}) = e_{ikl}\left(\underbrace{\epsilon^E_{jl,k}}_{\sum \square \approx 0} + \omega^E_{jl,k}\right) \qquad 5$$

$$\therefore \alpha_{ij} \cong \begin{bmatrix} \omega^E_{12,3} - \omega^E_{13,2} & \omega^E_{13,1} & \omega^E_{21,1} \\ \omega^E_{32,2} & \omega^E_{23,1} - \omega^E_{21,3} & \omega^E_{21,2} \\ \omega^E_{32,3} & \omega^E_{13,3} & \omega^E_{31,2} - \omega^E_{32,1} \end{bmatrix} \qquad 6$$

where $\beta$ is the local lattice distortion or displacement gradient tensor, $e_{kli}$ is the Levi–Civita permutation tensor, and $\epsilon_{jl,k}$ and $\omega_{jl,k}$ are the elastic strain and lattice rotation tensors, respectively.

With Nye's dislocation tensor ($\alpha$) written as the sum of the gradient of the lattice elastic strain and rotation (nine components); the EBSD technique cannot measure any derivatives normal to the sample surface, i.e., terms containing $\frac{\partial}{\partial x_3}$. In addition, experimentally, the elastic strain gradient was found to be sufficiently smaller than the rotation gradient (≈1%) [22,25,26];

therefore, it was neglected. This makes the five components of the dislocation tensors accessible ($\alpha_{13}, \alpha_{23}, \alpha_{33}, \alpha_{12}$ and $\alpha_{21}$), with the ($\alpha_{11} - \alpha_{22}$) being calculable since $\omega_{12,3}^E = -\omega_{21,3}^E$. These six components (nine in 3D techniques [2,27,28]) can be expressed as a column vector ($\Lambda$) for rotation derivatives, and are related to each slip system (D, with $b_i^d$ and $t_i^d$) using matrix (A) multiplied by the dislocation density matrix ($\rho$) with the matrix size in brackets (), as in equations (7) and (8), where the comma indicates partial differentiation with respect to Cartesian coordinates [25,29–31].

$$A\,(6*D).\,\rho\,(D*1) = \Lambda\,(6*1) \qquad 7$$

$$\begin{pmatrix} b_1^1 t_1^1 - \tfrac{1}{2}\sum_{i=1}^{3} b_i^1 t_i^1 & \cdots & b_1^d t_1^d - \tfrac{1}{2}\sum_{i=1}^{3} b_i^d t_i^d \\ b_1^1 t_2^1 & \cdots & b_1^d t_2^d \\ b_1^1 t_3^1 & \cdots & b_1^d t_3^d \\ b_2^1 t_1^1 & \cdots & b_2^d t_2^d \\ b_2^1 t_2^1 - \tfrac{1}{2}\sum_{i=1}^{3} b_i^1 t_i^1 & \cdots & b_2^d t_2^d - \tfrac{1}{2}\sum_{i=1}^{3} b_i^d t_i^d \\ b_2^1 t_3^1 & \cdots & b_2^d t_3^d \end{pmatrix} \begin{pmatrix} \rho^1 \\ \vdots \\ \rho^d \end{pmatrix} = \begin{pmatrix} \omega_{23,1} \\ \omega_{31,1} \\ \omega_{12,1} \\ \omega_{23,2} \\ \omega_{31,2} \\ \omega_{12,2} \end{pmatrix} \qquad 8$$

The least-squares method was used to solve for $\rho$. The solution to the minimisation problem of the dislocation density values is then obtained by equation 9, which is overdetermined when solving for any material except simple cubic materials; thus, a minimisation approach must be employed. That is either L$_1$, which is energetically motivated, or L$_2$, which is geometrically motivated, which is combined with a weighting factor (w) to reflect each system activation probability or (strain) energy [29].

$$\rho = (A^T A)^{-1} A^T \Lambda \qquad 9$$

Pantleon [29] argued that Nye's dislocation density tensor is the build-up/summation of the absolute values of single dislocations with different densities required to achieve the minimum total energy. Thus, the edge and screws were weighted (i.e. using a vector coefficient), as described in 10 and 11, based on their strain energy [32,33], where the effective Poisson ratio ($v$) was calculated from the Voigt stiffness tensor ($C_{ij}$) of an anisotropic material [34]. The GND density calculated with this method is termed a 'lower-bound GND.'

The L$_1$ method (natively implanted in MATLAB in the '*linprog*' algorithm) was employed (as in equation 10) because it has been widely used and shows high sensitivity in resolving the structure of individual dislocation variants [5,22,33,35]. In addition, this study assumed a direct relationship between the dislocation strain energy and dislocation length (through plastic strain gradients, dislocation density, and lattice curvature linear relations), irrespective of the negligible elastic strain gradients and dislocation field interaction influences, which are commonly used in the literature [5,19,25,29,36][2].

$$L_1 = \sum_{i=1}^{D} |\rho_d * w_d|, \qquad w_d^{screw} = (b_d^{screw})^2, \qquad w_i^{edge} = \frac{\left(b_d^{edge}\right)^2}{(1-v)} \qquad 10$$

$$v = \frac{C_{11} + 4C_{12} - 2C_{44}}{2(2C_{11} + 3C_{12} + C_{44})} \qquad 11$$

As tabulated in Table 1, Plastic deformation by dislocations in body-centred cubic ferrite crystals can be characterised by 28 characteristic edge or screw dislocation types, i.e., four <111> screw dislocations plus 24 edge dislocations of <$\bar{1}$11>{110} and <11$\bar{1}$> {112}, while <111> {123} dislocations were neglected because they are rarely activated [37].

Dislocation density estimation using HR-EBSD data can resolve the density of individual dislocation types [22,35,38]. As in [22], the analysis assumes pure characteristics of dislocations (edges or screws) and ignores partial and mixed dislocation characteristics. The Burgers and line vectors (equation 12) were transformed into the grain crystal frame (equation 13) before estimating their dislocation density [25].

---

[2] For L$_2$ method (equation below) can be calculated using Moore-Penrose pseudo-inversion (*pinv*) or nonnegative least-squares curve fitting problems (*lsqnonneg*).

$$L_2 = \sqrt{\sum_{i=1}^{D} (\rho_i * w_i)^2} \qquad 16$$

Table 1: Description and labelling of 28 distinct edge and screw dislocations for body-centred cubic (BCC) ferrite, where $b$ is the Burgers vector, and $t$ is the line vector (parallel to $b$ for screw dislocations). For dislocations with edge characteristics, $b$ is perpendicular to the dislocation line, and $t$ is a unit vector parallel to the dislocation line (i.e. $t \cdot b = 0$).

| Label | Edge $b$ | $t$ | Label | Edge $b$ | $t$ | Label | Edge $b$ | $t$ | Label | Screw $b$ |
|---|---|---|---|---|---|---|---|---|---|---|
| 1 | $[\bar{1}11]$ | $[\bar{1}1\bar{2}]$ | 9 | $[\bar{1}11]$ | $[211]$ | 17 | $[1\bar{1}1]$ | $[\bar{1}\bar{1}0]$ | 25 | $[1\bar{1}1]$ |
| 2 | $[1\bar{1}1]$ | $[\bar{1}12]$ | 10 | $[\bar{1}\bar{1}1]$ | $[\bar{2}11]$ | 18 | $[11\bar{1}]$ | $[0\bar{1}\bar{1}]$ | 26 | $[\bar{1}\bar{1}1]$ |
| 3 | $[1\bar{1}1]$ | $[\bar{2}\bar{1}1]$ | 11 | $[1\bar{1}1]$ | $[121]$ | 19 | $[11\bar{1}]$ | $[101]$ | 27 | $[\bar{1}11]$ |
| 4 | $[11\bar{1}]$ | $[2\bar{1}1]$ | 12 | $[\bar{1}\bar{1}\bar{1}]$ | $[1\bar{2}1]$ | 20 | $[\bar{1}11]$ | $[110]$ | 28 | $[11\bar{1}]$ |
| 5 | $[\bar{1}11]$ | $[12\bar{1}]$ | 13 | $[\bar{1}11]$ | $[01\bar{1}]$ | 21 | $[1\bar{1}1]$ | $[011]$ | | |
| 6 | $[11\bar{1}]$ | $[1\bar{2}1]$ | 14 | $[1\bar{1}1]$ | $[\bar{1}01]$ | 22 | $[\bar{1}\bar{1}\bar{1}]$ | $[0\bar{1}1]$ | | |
| 7 | $[11\bar{1}]$ | $[112]$ | 15 | $[11\bar{1}]$ | $[1\bar{1}0]$ | 23 | $[\bar{1}\bar{1}\bar{1}]$ | $[10\bar{1}]$ | | |
| 8 | $[\bar{1}\bar{1}\bar{1}]$ | $[11\bar{2}]$ | 16 | $[\bar{1}11]$ | $[\bar{1}0\bar{1}]$ | 24 | $[\bar{1}\bar{1}\bar{1}]$ | $[\bar{1}10]$ | | |

$$\hat{t} = \frac{t}{|t|}, \qquad \hat{b} = \frac{b}{|b|}a \qquad\qquad 12$$

$$t' = R_{\varphi_1, \Phi, \varphi_2}\begin{bmatrix}1 & 0 & 0\\ 0 & 1 & 0\\ 0 & 0 & 1\end{bmatrix}\hat{t}, \qquad b' = R_{\varphi_1, \Phi, \varphi_2}\begin{bmatrix}1 & 0 & 0\\ 0 & 1 & 0\\ 0 & 0 & 1\end{bmatrix}\hat{b} \qquad 13$$

The normal direction (ND), parallel to $x_3$, of the plane $(hkl)$ that is parallel to the sample surface was estimated from the mean grain orientation. The angle ($\Phi$) between the Burgers vector of each dislocation type and normal direction was obtained using equation 14. In the analysis of the dislocation density at the slip bands, the orientations ($\Phi_i$) at each of the measured points in the slip band were weighted by the relative contribution of each dislocation type to the total GND density to calculate the weighted average angle ($\bar{\Phi}$, Figure 1) for the dislocations (i.e. the dot product of both $b$ and $t$ with the slip plane normal should equal zero)[32].[3]

$$\Phi_i = \tan^{-1}\left(\frac{\|b'_i \times (hkl)\|}{b'_i \cdot (hkl)}\right) \qquad 14$$

---

[3] A function with an example is available at https://doi.org/10.5281/zenodo.6411623.

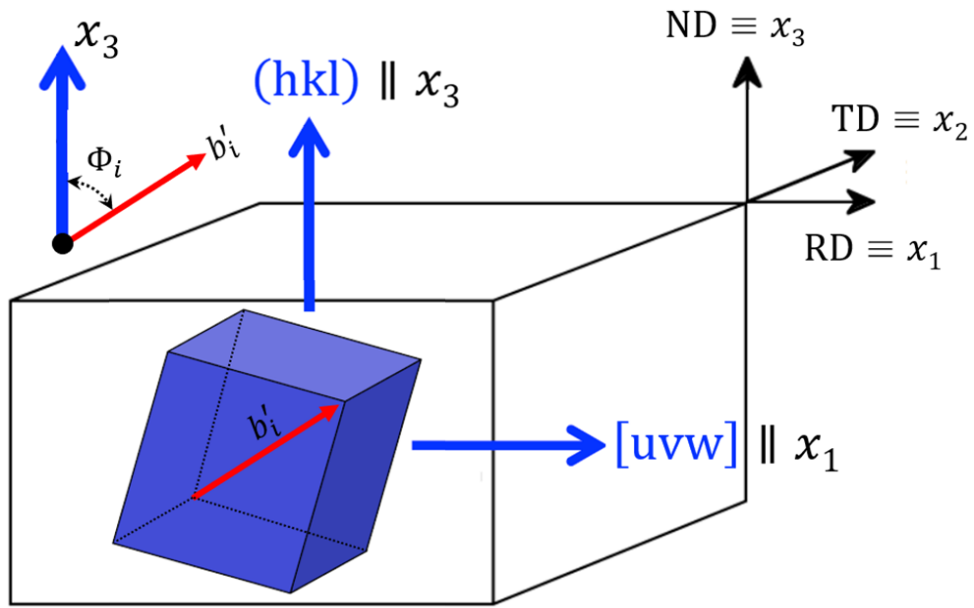

Figure 1: Schematic of $\Phi_i$ calculation showing an arbitrarily oriented crystal (blue) in the sample (reproduced from [39]).

# 3. Results

Some slip bands were found to terminate at very low-angle (sub) grain boundaries (<3° grain boundary misorientation). In contrast, others terminated mid-grain with no indication of the existence of an obstacle, either on the surface or after section using focused ion beam (FIB). However, no observations were made of slip-band piecewise propagation (while maintaining both the direction and plane) with the load. A closer look was made at the slip-band deformation field during the early stages of inception, loaded stationary slip bands, and slip-band arrays. However, due to high magnification and slow pattern acquisition, rapid carbon deposition accumulation introduced noise to the measurement and frequently limited additional acquisition at the same location after further sample straining.

The average measurement mean angular error was $2.7 \times 10^{-4}$, which makes the sensitivity limit of the estimated total GND density approximately $1.45 \times 10^{13}$ m$^{-2}$, as calculated using the Wilkinson and Randman [25] equation to estimate the dislocation density measurement sensitivity limit.

## 3.1. Incipient slip band

A slip band was formed within the ferrite grain, exhibiting a small step-like feature (Figure 2a). Using EBSD analysis [40], the orientation of the grain was examined, revealing that the slip band followed the trace of the (112) slip plane, which was inclined at approximately 90° to the surface, as validated using focused ion beam (FIB) milling.

The GND analysis of the incipient slip band showed a higher edge dislocation activity than that of the screw dislocations (Figure 2b). By weighting the relative contribution of the identified dislocation systems (Figure 2c), $\bar{\Phi}$ was 50.9 ± 1.2°; and the variance was obtained from four repeated measurements (within the blue-shaded region in Figure 2b). Although this direction indicates significant out-of-surface shear and shear parallel to the surface; none of the slip systems identified in the GND analysis (Table 1) are mobile on the (112) plane, i.e., the equivalent Burgers vector is not in the plane of the slip band. This suggests that the probed dislocations accommodate the slip band's associated deformation, particularly the complex local lattice distortion near the slip band height/step.

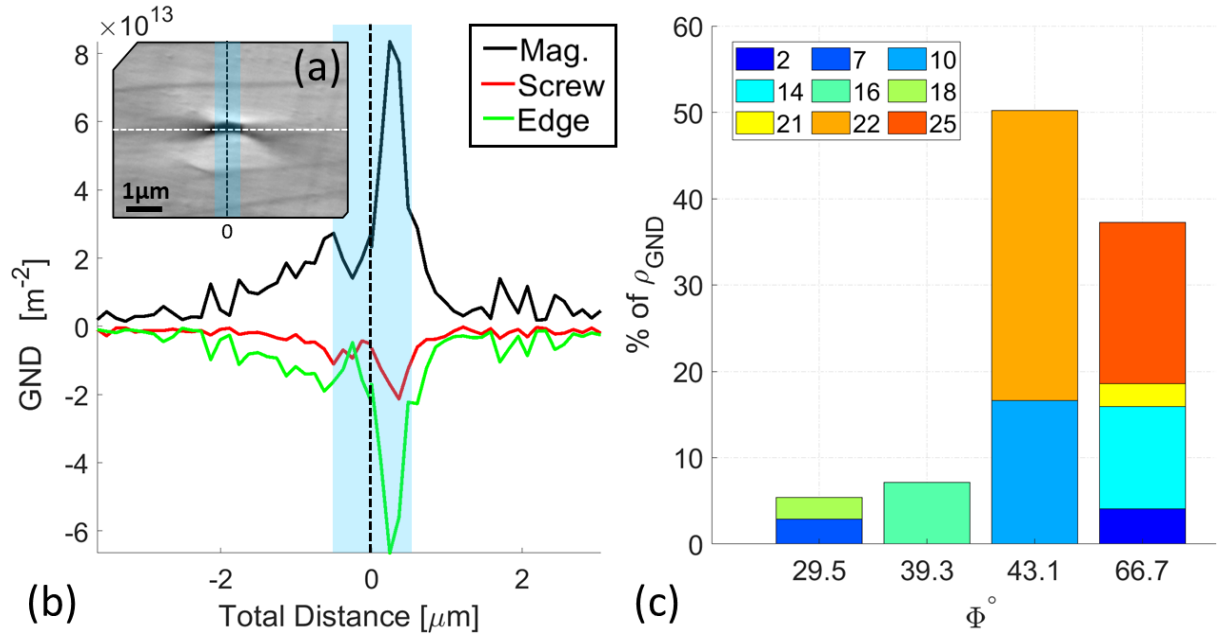

Figure 2: (a) Secondary electron image of the incipient slip band. (b) Dislocation activity across the $[11\bar{1}](112)$ slip band. The dislocation-type sign indicates the local misorientation polarity, with negative values representing depression-like (concave) activities and positive values indicating what is commonly known as "pile-ups" (convex) activities [35]. (c) Angle ($\Phi$) of each dislocation system of the Burgers vector relative to the surface, with the relative contribution of each system stacked at the related angle. The dislocations forming the slip band were obtained from the area shaded in blue in (a) and (b). The dislocation systems associated with each label are listed in Table 1 (reproduced from [39]).

## 3.2. Array of slip bands

An array of slip bands was observed within the ferrite grain. At the tip of each slip band, a 'blooming[4] zone' of diffused slips was observed (Figure 3a). The average slip bands thickness were 0.59 ± 0.02 µm, 0.61 ± 0.02 µm, and 0.44 ± 0.01 µm, respectively, measured from the secondary electron (SE) image and GND density map. The surface trace analysis identified the array of slip bands as $[1\bar{1}1](\bar{1}12)$ slip bands with a 107.6 ± 9.5° inclination angle ($\psi$) of the Burgers vector from the surface trace. $\psi$ is the angle between the slip direction and the observed trace [41], which indicates that a significant out-of-plane shear occurred.

The GND density, presented as a function of the distance along the slip band and extending beyond its tip into the parent grain, is presented for each slip band in Figure 3a. There is a sharp drop in the GND dislocation activity outside the slip band, i.e., the blooming zone, and the GND density appears to be higher near the tip of the narrower slip band. Analysis of the individual dislocation types using the estimated GND density inside the slip band revealed that all dislocation types were active, particularly edge dislocations. The weighted average angle, $\overline{\Phi}$, between the Burgers vector and the surface normal for all dislocation types was 46.1 ± 2.4° (the variance was obtained with four repeated measurements for each slip band, i.e., 12 in total).

---

[4] "Blooming," "avalanche," or "dislocation burst" are used in this article interchangeably.

However, when only considering the three dislocation types (two edges and one screw) – that were mobile on $(\overline{1}12)$, i.e., $[1\overline{1}1]b\ [\overline{1}\overline{1}0]t$, $[1\overline{1}1]b\ [110]t$, and $[1\overline{1}1]b\ [1\overline{1}1]t$, which comprised 2.4%, 3.0%, and 12.0% of all dislocations measured in the slip band, respectively – the weighted Burgers vector ($\overline{\Phi}$) of the mobile dislocations was 12.7° relative to the surface normal. These are dislocations of the $[1\overline{1}1](\overline{1}12)$ system with the highest Schmid factor (i.e. with a slip inclination angle, $\psi$, of 107.6°); however, it is also clear that most of the GND density was due to the accommodation of other deformations.

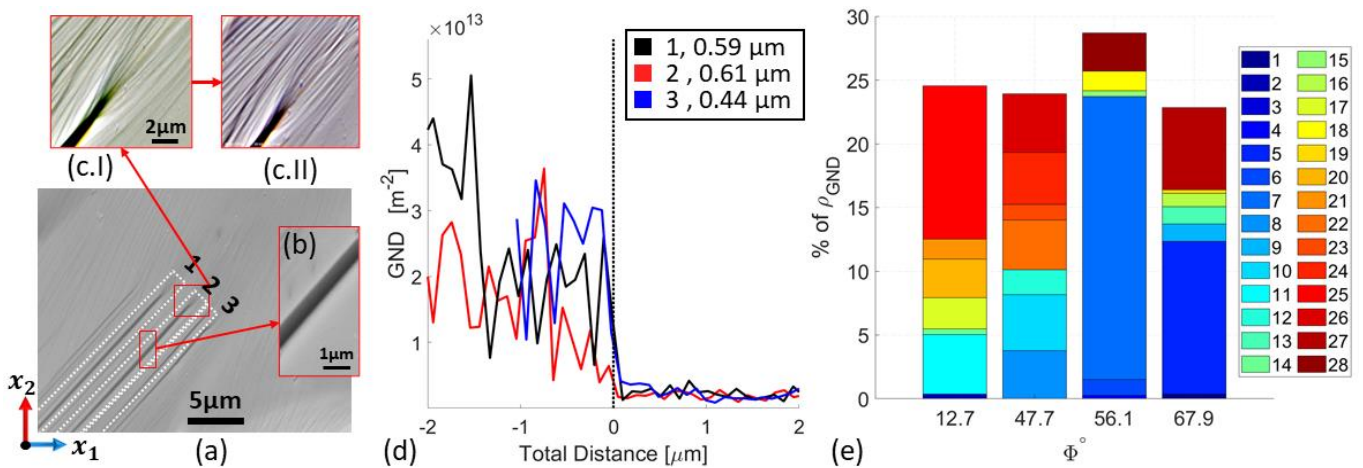

Figure 3: An array of $[1\overline{1}1](\overline{1}12)$ slip bands within a ferrite grain; (a) SE image with (b) zoomed image of slip band no. 2; (c) Secondary electron (SE) of the tip of slip band no.2 at applied displacements of (I) 1.1 mm and (II) 1.25 mm. a-c are reproduced from [39]. (d) Dislocation activity across three slip bands in the $(\overline{1}12)$ slip band array (0 is at the visually identified slip band tip in each case). The legend includes the slip band thickness as measured from the SE images. (e) Angle (Φ) of each dislocation type for the Burgers vector relative to the surface, averaged over the distance (-2 to 0) for the three slip bands. The dislocation types associated with each label are listed in Table 1.

## 3.3. Loading and unloading of a slip band

The impact of the load variation on the intragranular ferrite slip band in the middle of the grain was examined through in situ observations (Figure 4a). This slip band intersected a low-angle grain boundary (represented by a dashed white line), and its trace was aligned with the (112) plane inclined at approximately 42° to the surface. Another secondary slip band with a trace corresponding to (011) was also present in the field of view. Utilising the local loading direction matrix, it was determined that the $[11\bar{1}](112)$ slip system exhibited the highest Schmid factor, followed by $[11\bar{1}](011)$. The shear direction in the $[11\bar{1}](112)$ slip system had an inclination angle ($\psi$) of 172°, indicating that the Burgers vector was nearly parallel to the observed surface.

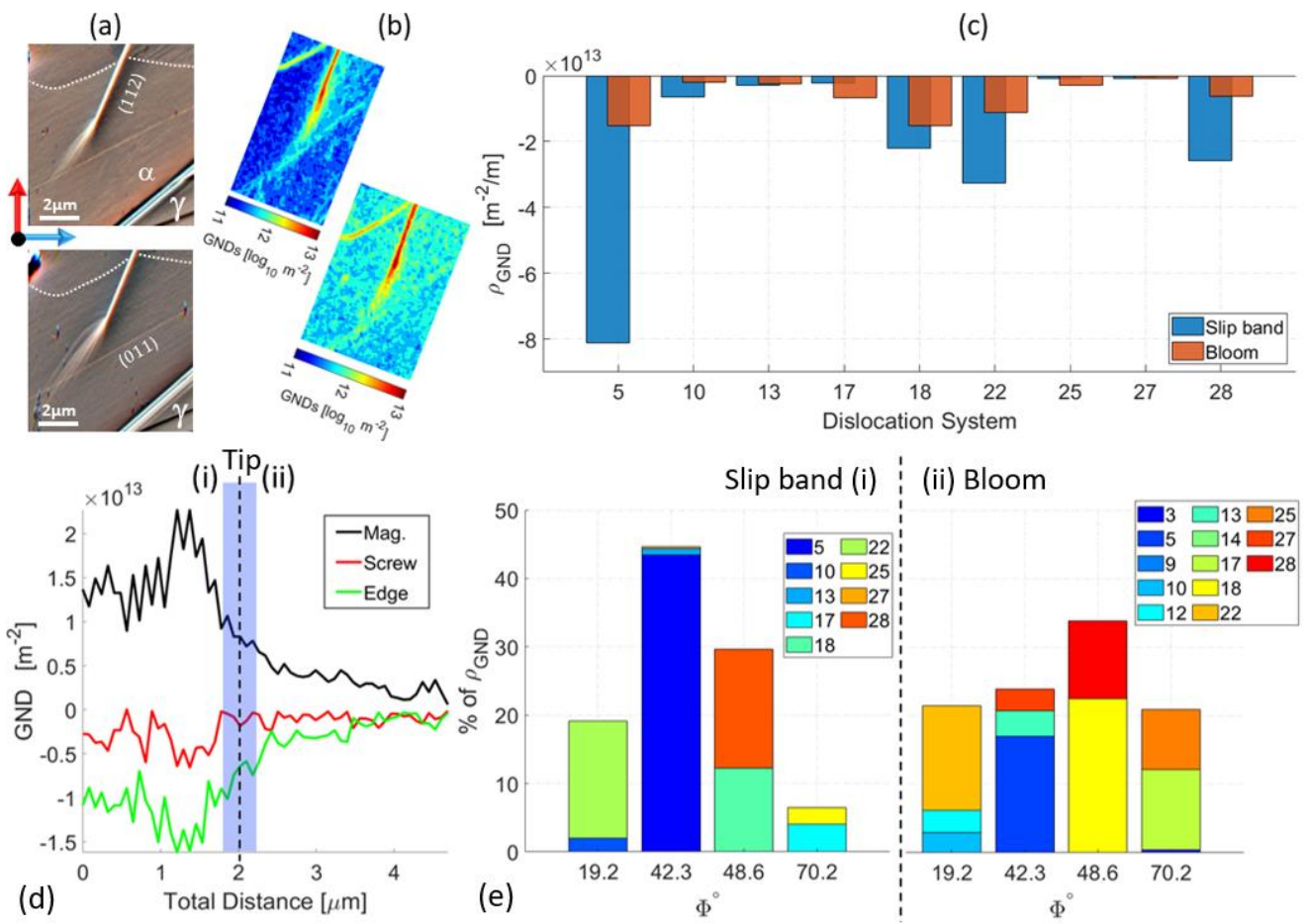

Figure 4: In situ loading of a $[11\bar{1}](112)$ slip band in a ferrite (a) grain observed using (a) VFSD imaging and (b) GND density. Notations (i) and (ii) indicate observations with (i) 1 mm and (ii) 1.2 mm extensions. (d) Dislocation activity in $[11\bar{1}](112)$ slip band at an applied displacement of 1 mm. (e) The angle ($\Phi$) of each dislocation system's Burgers vector to the surface normal with their relative contribution, inside the slip band (the area from 0 until the dashed line near the 2 μm marker; (c) The dislocation activities in the slip band and the blooming zone. The dislocation system associated with each label is shown in Table 1. The sign of the GND density indicates the local misorientation direction, with negative values representing concave-like activities and positive pile-up activities [35] (reproduced from [39]).

As the applied load increased, the contrast variation at the tip of the slip band due to dislocation bursting was expanded. However, no measurable changes were observed in the thickness (height) of the slip band, which remained constant at 0.21 µm. The total GND density, estimated by averaging within the cropped field of view encompassing the dislocations of the slip band, increased slightly with the applied displacement. At 1 mm, the GND density was measured to be 11.96 ± 0.03 log($m^{-2}$), while at 1.2 mm, it was 12.15 ± 0.02 log($m^{-2}$).

The dislocation types in the slip band and at the tip vicinity were characterised using the example at an applied sample displacement of 1 mm. The dislocation activity was highest inside the slip band but was also significant in the bloom zone ahead of the tip. The slip band had high edge dislocation activity (79.8%), especially $[\bar{1}11]b\ [112]t$ which contributed to 43.4% of the total GND density. The weighted average angle ($\bar{\Phi}$) to the surface normal was estimated to be 41.6 ± 1.7°. However, the $[11\bar{1}]b\ [11\bar{1}]t$ screw was the only mobile dislocation type on the slip band plane. As noted above, the inclination angle ($\psi$) of the $[11\bar{1}]$ Burgers vector was 172.3° to the slip trace and 48.6° to the surface normal. These mobile screw dislocations comprised only 17.4% of the total dislocations that formed the slip band. The blooming zone had a much lower dislocation activity, especially the $[\bar{1}11]b\ [12\bar{1}]t$ edge dislocations.

## 4. Discussion

The observed slip bands were found to be terminated within the grain interior, seemingly unaffected by any discernible obstacles, which was further substantiated using focused ion beam (FIB) milling, as reported in [11]. One reason for this is the strain gradient from the 3-point bend test, as in this study, the investigated grains were between the tension and the stress' neutral axis. Thus, the propagation of the slip band either increased with load or hampered by the opposing deformation gradient, given the large ferrite grain size (around 0.25 mm$^2$, see the Supplementary Figures). In addition, a contributing factor to the slip band termination mid-grain comes from the aging of duplex stainless steel (DSS). DSS alloys can have limited toughness due to their large ferritic grain size, and the ferrite ($\alpha$) hardening and embrittlement tendencies at temperatures between 280–500 °C, especially at 475 °C. This is known as "475 °C embrittlement", and at this temperature range, spinodal decomposition of ferrite phase into Fe-rich nanophase ($\alpha'$) and Cr-rich nanophase ($\alpha''$), accompanied by G-phase precipitation occurs [13,42,43]. Spinodal decomposition increases the hardening of ferrite owing to the misfit between the Cr-rich and Fe-rich nano-phases, variations in the elastic modulus, and internal stress [43–45], which impedes dislocation movement [46,47]. This, in turn, affects the nucleation and propagation of slip bands in the ferrite matrix.

Thus, the created pencil slip band does not get thick and remains planar owing to the spinodal nanostructure obstructing dislocation movement [48,49]. Simultaneously, the external stress intensifies the tip-generated stress field (compared to the matrix) and dislocations, preparing the slip to propagate [50,51]. In the cases observed here, the tip-generated dislocations have non-planer cores and will cross-slip (without dissociating) between the many available slip planes, following the easiest path creating the 'wavy slips' that was observed at the initial stage of deformation and in the blooming zones at the slip band tip (Figure 5).

The change in surface topography resulting from this dislocation burst is observed as the "blooming" contrast shown in Figure 5. The intensity of this topographical blooming and the magnitude of the elastic displacement increased with loading (Figure 4a) and the height of the slip band steps (Figure 5$\alpha_3$). To understand these blooming zones, we need to consider that in general, a reduction in the mobile dislocation density may arise from dislocations escaping from free surfaces, dislocation source shutdown, and mechanical annealing [3,52]. Du et al. [50] revealed that the visible burst traces observed on the surface, similar to those depicted in Figure 4a and Figure 5, were a result of the dislocation loops reaching the surface and serving as a stress-relieving mechanism. When the stress at the dislocation sources reaches a critical value, new avalanches occur as the previous avalanche moves perpendicular to the slip plane. The generation of new avalanches is hindered by back stress, which can be overcome by increasing the external stress [49]. Consequently, this leads to intermittent periods of internal short blooms and occasional large surface blooms or bursts at the slip band owing to stress accumulation and the repulsion of immobile dislocations near the tip region [50,53,54].

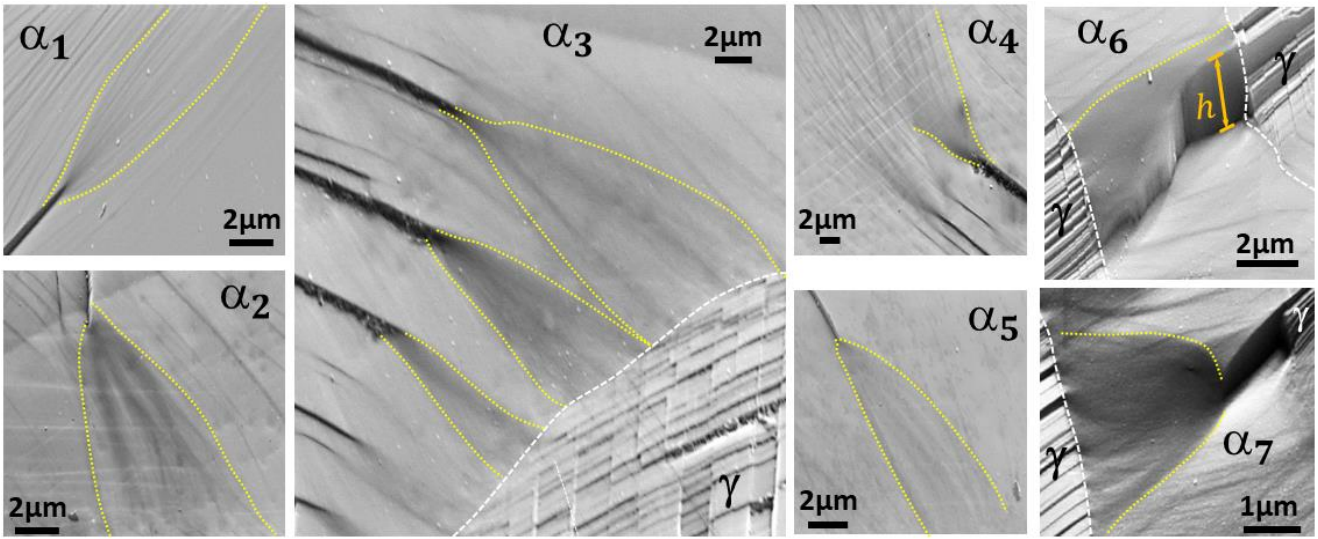

Figure 5: In-lens images of dislocations blooming from the tip of slip bands in the ferrite phase of age-hardened duplex stainless steel. $h$ in $\alpha_6$ denotes the slip band step height.

Eventually, as deformation progresses, the applied shear forces dislocations to nearly planar rearrangements (assisted by thermal activation) before plane-glide into a pencil-like shape [55], as observed in the incipient slip band. Although this arrangement is due to shear, the plane of choice does not follow Schmid's law and is affected by tension-compression asymmetry and non-shear stresses [56]. The shear and dilatational stress components arise from the fields around the spinodal precipitates owing to loop accumulation [57]. Therefore, slip bands represent localised sites of shear deformation caused by mobile dislocations and the observed dislocation burst at the tip of the slip band alleviates the pressure stress field [49,58], when the surrounding matrix yields near the slip band tip [59].

From the presented data and previous work on the elastic deformation field [11], the observed slip bands indicate a highly localised shear deformation with edge dislocations dominating within the band to accommodate the deformation caused by shear and complex strains at the surface of the specimen. This is because, in age-hardened stainless steel, the created pencil slip band tends to remain planar as the spinodal nanostructure obstructs dislocation movement [48]. Furthermore, by resolving the individual dislocation types at the location of the pencil-like slip band, we found that the dislocation activity was mainly due to immobile dislocations on the slip band plane. These immobile dislocations are necessary to maintain the geometrical distortion of the slip band. A higher dislocation activity was observed at the slip band with more edge dislocation activity compared to screw dislocations, which is associated with high GND density (or hot) spots and a lower slip band height (Figure 3a) in the array of slip bands. This is consistent with previous observations that found that the thickness of the stress concentration layer is inversely related to the dislocation density [60]. This length-scale aspect resembles the low GND density around large indents [61]. The GND density decreased abruptly ahead of the slip band tip, similar to that observed at the crack tips, and was linked to an intense strain gradient [62]. Therefore, the GND density is related to the plastic strain gradient that locally accommodates the slip band shear.

This interplay between the shear stress and dislocation can be better understood by looking at the EBSD-probed deformation field (Figure 6a and b) for $\alpha_6$ slip band that was shown Figure 5. The HR-EBSD deformation field around $\alpha_6$ slip band the shows that inside the slip band, GND activity is very high; however, outside the slip band, at the blooming zone, there is high shear stress with very low GND density, especially in mobile dislocation density, which can be due to dislocations escaping from free surfaces, dislocation source shutdown, and mechanical annealing [3,52], as discussed earlier. The in-plane shear stress map outside the $\alpha_6$ slip band matches the blooming zone that is extenuated by the yellow line in Figure 5, especially the blooming's bottom edge. Thus, this indicates that the blooming zone associated with intragranular slip bands is not directly correlated with a high GND density or changes in lattice rotation, which contrasts with the blooming observed ahead of the slip bands impinged at grain boundaries [63–66]. However, it exhibits similarities to the surface displacement bursts caused by the internal movement of dislocations observed in nano-compression tests [53,54,67].

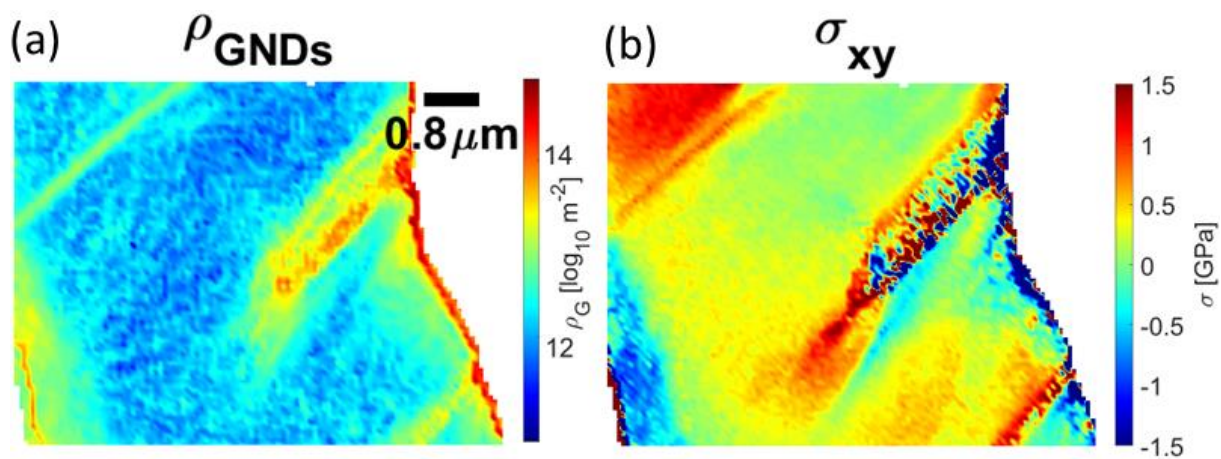

Figure 6: HR-EBSD (a) GND density and (b) in-plane shear stress field resolved on the $[\bar{1}11](101)$ slip band ($\alpha_6$ slip band in Figure 5).

Nevertheless, it is critical to note that the current analysis hinges on the fact that the determination of Nye's tensor for lower-bound GND density is an estimation, that will benefit greatly from further studies that compare the measurement of dislocation density using transmission electron microscopy [68] with HR-EBSD estimation of the density of individual dislocations are needed to properly assess the propagation of uncertainty/error due to (1) EBSP's resolution [69], (2) EBSPs binning [70], (3) measurement step size [36,70], (4) HR-EBSD precision [10,25], and (5) GND density estimation methods.

## 5. Conclusions

This study investigated slip bands in ferrite grains under different conditions by resolving the individual dislocation types at the slip band location using the mapped geometrically necessary dislocation (GND) density. We found that edge dislocations were more active and predominant than screw dislocations in the slip band, and that the dislocation activity was mainly from dislocations that were immobile on the slip band plane and necessary to maintain the geometrical distortion of the slip band.

In addition, an inverse size effect was observed between the slip bands and the dislocation density. A 'blooming zone' at the slip band tip was associated with high-GND density (or hot) spots, which indicates a reduction in mobile dislocation density due to dislocation escape from the surface. This zone is linked to the intense strain gradient as these slip bands exhibited significant shear. The 'blooming zone' at the slip band tip increased in size with external load. In contrast, the thickness of the slip band remained unchanged.

# Acknowledgements

The author appreciates discussions with Professor James Marrow and Dr Phani Karamched (University of Oxford). The author would also like to thank Ms. Marzena Tkaczyk (Laboratory for In-situ Microscopy and Analysis, LIMA) for supporting the experiments. The author acknowledges the use of characterisation facilities within the David Cockayne Centre for Electron Microscopy (DCCEM), Department of Materials, University of Oxford, along with the financial support provided by the EPSRC (Grant ref. EP/N509711/1).

# CRediT author statement

**Abdalrhaman Koko:** Conceptualisation, Methodology, Visualisation, Investigation, Formal analysis, Writing - original draft

binning and step size. Ultramicroscopy 2013;125:1–9. https://doi.org/10.1016/J.ULTRAMIC.2012.11.003.